\documentclass[aps,superscriptaddress,prd,twocolumn,nofootinbib]{revtex4}

\usepackage{amsmath,bm}
\usepackage{graphicx}
\usepackage{epstopdf}
\usepackage{subfigure}
\usepackage{epsfig}
\usepackage{amsmath,amssymb,amsfonts}
\usepackage{multirow}
\usepackage{tabularx}
\usepackage{color}
\usepackage{palatino}
\usepackage[breaklinks,colorlinks,urlcolor=blue,citecolor=blue,linkcolor=blue]{hyperref}
\usepackage[utf8]{inputenc}
\usepackage{verbatim}
\usepackage{array}
\usepackage{marvosym}
\usepackage{float}
\def\bea#1\eea{\begin{align}#1\end{align}}

\newcommand{\bef}{\begin{figure}[!t]\centering}
\newcommand{\eef}{\end{figure}}
\newcommand{\eq}[1]{Eq.~(\ref{#1})}
\newcommand{\fig}[1]{Fig.~\ref{#1}}

\newcommand{\pT}{p_{\rm T}}
\newcommand{\pTD}{p_{\rm T,D^0}}
\newcommand{\pTJ}{p_{\rm T,jet}^{\rm ch}}

\graphicspath{{figures/}} 

\begin{document}
\title{Medium modifications of heavy-flavor jet angularities in high-energy nuclear collisions}

\date{\today  \hspace{1ex}}

\author{Yao Li}
\affiliation{Key Laboratory of Quark \& Lepton Physics (MOE) and Institute of Particle Physics, Central China Normal University, Wuhan 430079, China}

\author{Shi-Yong Chen}
\affiliation{Huanggang Normal University, Huanggang 438000, China}

\author{Wei-Xi Kong}
\affiliation{Key Laboratory of Quark \& Lepton Physics (MOE) and Institute of Particle Physics, Central China Normal University, Wuhan 430079, China}

\author{Sa Wang}
\email{wangsa@ctgu.edu.cn}
\affiliation{College of Science, China Three Gorges University, Yichang 443002, China}
\affiliation{Center for Astronomy and Space Sciences and Institute of Modern Physics, China Three Gorges University, Yichang 443002, China}

\author{Ben-Wei Zhang}
\email{bwzhang@mail.ccnu.edu.cn}
\affiliation{Key Laboratory of Quark \& Lepton Physics (MOE) and Institute of Particle Physics, Central China Normal University, Wuhan 430079, China}

\begin{abstract}
We present the first theoretical study of heavy-flavor jet angularities ($\lambda^{\kappa}_{\alpha}$) in Pb+Pb collisions at $\sqrt{s_{\rm NN}}=$ 5.02 TeV. The initial production of heavy-flavor jets is carried out using the POWHEG+PYTHIA8 prescription, while the jet evolution in the quark-gluon plasma (QGP) is described by the SHELL transport model. In p+p collisions, we observe narrower angularity distributions for D$^0$-tagged jets compared to inclusive jets, consistent with the ALICE preliminary results. We then demonstrate that jet quenching in the QGP slightly widens the angularity distribution of D$^0$-tagged jets in Pb+Pb collisions relative to that in p+p collisions for jet transverse momentum of $10 < \pTJ < 20$ GeV/c, while the angularity distributions of inclusive and D$^0$-tagged jets become narrower in Pb+Pb collisions relative to p+p at $\pTJ > 20$ GeV/c due to the strong influence of the selection bias. Additionally, by comparing the average angularities $\langle \lambda^{\kappa}_{\alpha} \rangle$ of inclusive, D$^0$-tagged and B$^0$-tagged jets with varying $\alpha$ and $\kappa$, we show that the larger the quark mass is, the lower the jet's $\langle \lambda^{\kappa}_{\alpha} \rangle$ values are. As a result of the slenderer initial distribution, we predict that as compared to inclusive jets, the heavy-flavor jets, especially the B$^0$-tagged ones, will suffer stronger modifications of $\langle \lambda^{\kappa}_{\alpha} \rangle$ in Pb+Pb relative to p+p at $10 < \pTJ < 20$ GeV/c. For a larger jet radius, a more significant broadening of jet angularities is predicted because of the enhanced contributions of the wide-angle particles.
\end{abstract}

\maketitle

\section{Introduction}
\label{sec-int}

One of the most important topics in heavy-ion physics is to understand the properties of the de-confined nuclear matter, the quark-gluon plasma (QGP), in high temperature and density environment \cite{Freedman:1976ub, Shuryak:1977ut}. When an energetic parton passes through the QGP created in high-energy nucleus-nucleus collisions, it may strongly interact with the thermal dense medium, resulting in a loss of its energy and broadening of its transverse momentum, referred to as the ``jet quenching'' effect \cite{Wang:1992qdg, Gyulassy:2003mc, Wang:2002ri, Vitev:2009rd, Neufeld:2010fj, Vitev:2008rz, He:2020iow, Chen:2022kic, Zhao:2021vmu, Yang:2023dwc, JETSCAPE:2022jer, Luo:2023nsi, Yang:2022nei, Zhang:2023oid, Zhang:2021xib, Xie:2024xbn, Chen:2024cgx, Xu:2014tda, Ma:2023zfj, Wu:2022vbu, Yan:2020zrz}. In particular, heavy quarks (charm and bottom) are powerful hard probes for studying the jet quenching effects because they are produced in the early stage of nuclear collisions and conserve their identity throughout the medium evolution; see reviews in Refs.~\cite{Dong:2019unq, Zhao:2020jqu, Tang:2020ame, Wang:2023eer}. By studying the production of heavy-flavor mesons and jets (a spray of collimated hadrons) in high-energy heavy-ion collisions, one can effectively extract the transport properties of the QGP and further test the quantum chromodynamics (QCD) under extreme conditions \cite{Francis:2015daa, Xu:2017obm, Cao:2018ews, Li:2019lex, Kumar:2020wvb, JETSCAPE:2022hcb}.

In the last two decades, experimental physicists have made great efforts to investigate the behavior of heavy quarks in heavy-ion collisions, including yield suppression \cite{CMS:2017qjw, PHENIX:2011img, STAR:2013eve, ALICE:2014wnc}, collective flow \cite{Adamczyk:2017xur, Sirunyan:2017plt, STAR:2019clv}, and baryon-to-meson ratio of heavy-flavor hadrons \cite{STAR:2019ank, Vermunt:2019ecg}. At the same time, theorists have conducted extensive and in-depth research on heavy quark production \cite{Cacciari:2005rk, Kniehl:2004fy}, cold nuclear matter (CNM) effect \cite{Eskola:2009uj, Eskola:2016oht}, mass-dependent energy loss \cite{Cao:2016gvr, Li:2020umn, Xing:2023ciw} and hadronization mechanism \cite{Cao:2019iqs, Plumari:2017ntm, He:2019vgs}. Recently, studies on the heavy-flavor jet observables, such as yield suppression \cite{Li:2018xuv, Kang:2018wrs}, transverse momentum balance \cite{Dai:2018mhw, Li:2024uzk} and azimuthal correlations \cite{Wang:2020qwe}, provide additional insights for
further understanding the mass and flavor dependence of energy loss. In particular, with the help of the increased statistics and improvement in experimental detection accuracy, the substructure of heavy quark jets, such as jet radial profile \cite{Wang:2019xey, Wang:2020ukj}, jet fragmentation function \cite{Li:2022tcr} and splitting function \cite{Li:2017wwc}, has gradually become a hot topic of theoretical and experimental researches in recent years. Substructure observables describe a jet's internal energy distribution and splitting configuration \cite{Ringer:2019rfk, JETSCAPE:2023hqn, Milhano:2017nzm, Caucal:2019uvr,Wang:2022yrp}, thus providing a unique perspective to explore interactions between heavy quarks and the
QGP in detail \cite{Wang:2019xey, Li:2017wwc, ALICE:2022phr}.

The generalized jet angularities, quantifying the transverse momentum ($\pT$) and angular distributions of constituents within the jet, form a class of jet substructure observables \cite{Larkoski:2014pca}, which are defined as
\bea
\lambda^{\kappa}_{\alpha} = \sum_{i {\rm \in jet}}
\bigg( \frac{p_{{\rm T},i}}{p_{\rm{T,jet}}^{\rm ch}} \bigg)^{\kappa}
\bigg( \frac{\Delta R_{{\rm jet},i}}{R} \bigg)^{\alpha}
\equiv \sum_{i {\rm \in jet}} z_i^\kappa
\theta_i^{\alpha},
\label{eq:lambda}
\eea
where the index $i$ sums over all the jet constituents, $R$ is the jet radius, $\Delta R_{{\rm jet},i} = \sqrt{(y_{\rm jet} - y_i)^2 + (\phi_{\rm jet} - \phi_i)^2}$ is the angular distance between jet constituent $i$ and the jet axis in the rapidity
($y$) and azimuthal angle ($\phi$) plane. The parameters $\kappa$ and $\alpha$ control the weights of momentum fraction and opening angle of jet constituents. Recently, the ALICE collaboration has reported the measurement of D$^0$-tagged jet angularities in p+p collisions at $\sqrt{s}=5.02$ TeV \cite{Dhankher:2024rkv}, which show smaller values compared to the inclusive jets at the same kinematic region. Since the angularity quantifies the $\pT$-weighted angular distribution of particles in jets \cite{Larkoski:2014pca}, the smaller angularity of D$^0$-tagged jets indicates more concentrated energy distribution around the jet axis compared to inclusive jets. It is of great importance to reveal the nature of different substructures between D$^0$-tagged jets and inclusive jets \cite{Chien:2024uax, Caletti:2021oor, ALICE:2021njq, Reichelt:2021svh, Budhraja:2023rgo}. Compared to previous heavy-flavor jet substructure measurements which focus on either the angular distribution (the radial profile) or momentum fractions (the fragmentation function and splitting function) of jet constituents, jet angularities allow us to study the interplay between the two as the heavy quarks traverse the QGP \cite{Larkoski:2014pca, Yan:2020zrz}. Therefore, it is of theoretical significance to investigate the heavy-flavor jet angularities in Pb+Pb collisions, which can be tested by future measurements at the LHC.

We present theoretical studies of heavy-flavor jet angularities in p+p and Pb+Pb collisions at $\sqrt{s_{\rm NN}}=5.02$ TeV. We will discuss differences in angularities between inclusive and D$^0$-tagged jets. In nucleus-nucleus collisions, we will present first calculations of heavy-flavor jet angularities and compare them with those of light-flavor jets. We further test the sensitivities of jet angularities and their modifications to the parameters $\alpha$ and $\kappa$.

\section{Jet angularities in p+p collisions}
\label{sec-pp}

\begin{figure}[htbp]
  \includegraphics[width=0.85\linewidth]{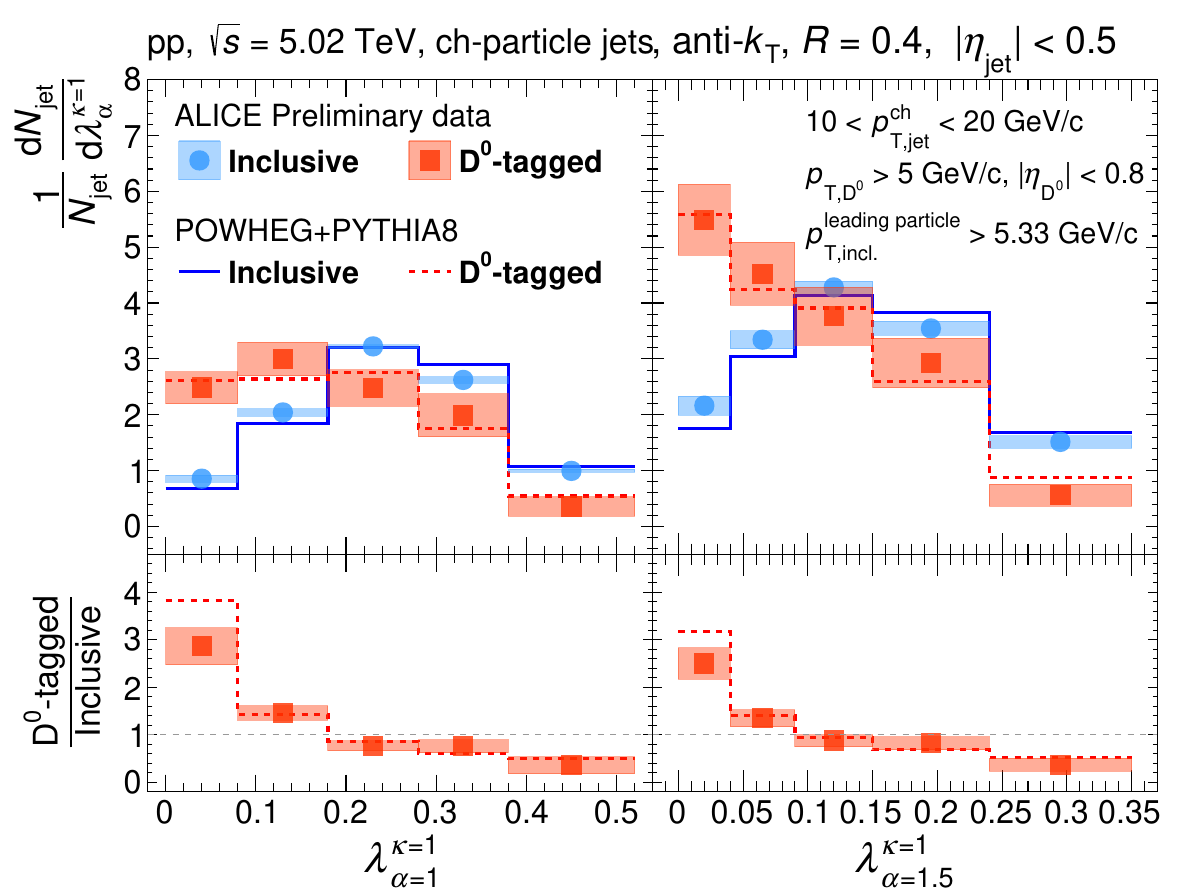}
  \includegraphics[width=0.85\linewidth]{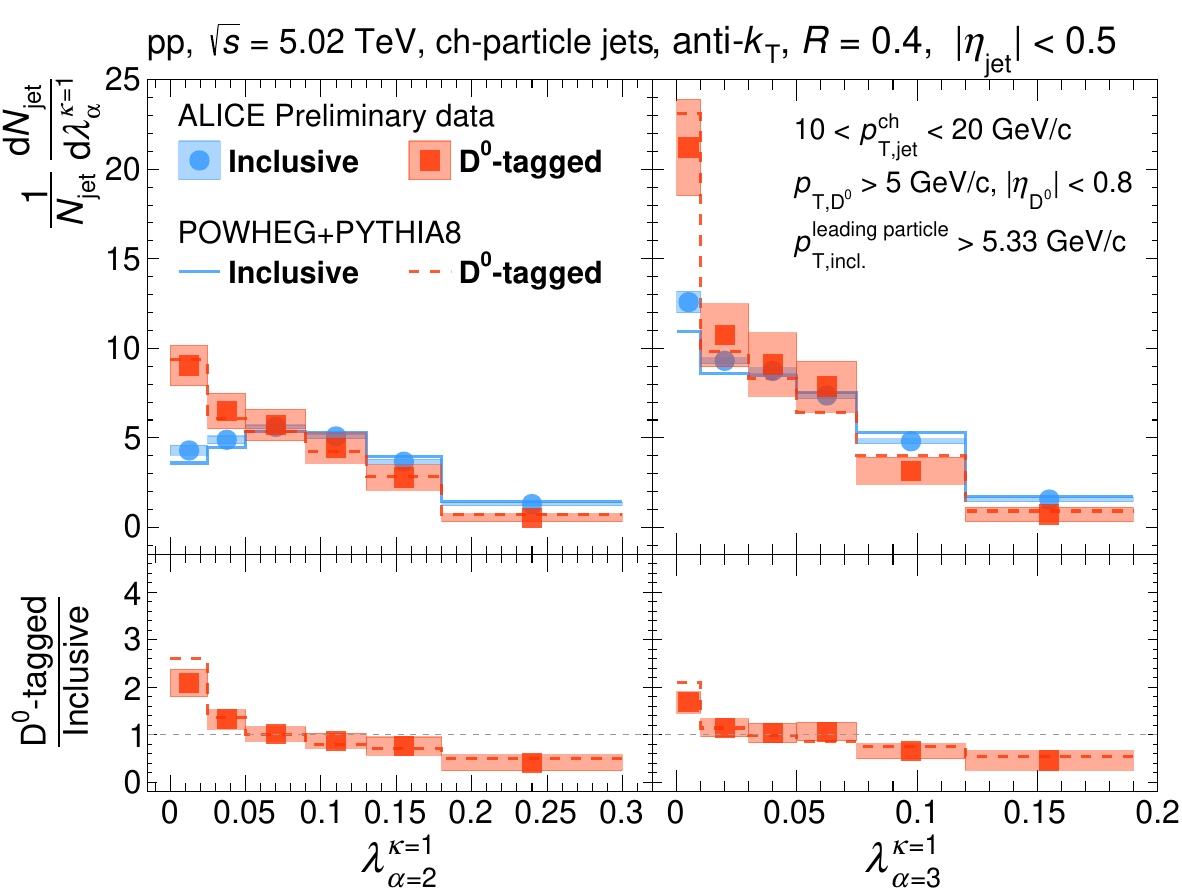}
  \caption{ (Color online) Normalized angularity distributions of D$^0$-tagged and inclusive jets for four different values of $\alpha$: 1.0, 1.5, 2.0, and 3.0 with the fixed $\kappa=$ 1.0 in p+p collisions at $\sqrt{s}=5.02$ TeV, compared to the ALICE preliminary data~\cite{Dhankher:2024rkv}.}
  \label{fig:D0_baseline}
\end{figure}

In this work, we simulate the production of inclusive and heavy-flavor jets with the POWHEG+PYTHIA8 prescription which matches the next-to-leading order (NLO) matrix elements with the resummation of parton shower (PS)~\cite{Frixione:2007vw, Alioli:2010xd, Bierlich:2022pfr}. The QCD di-jet processes \cite{Alioli:2010xa} and CT18NLO parton distribution function (PDF)~\cite{Hou:2019qau} are used in the simulations. The hadronization for partons in vacuum is simulated by the Lund string model implemented in PYTHIA8~\cite{Bierlich:2022pfr}. Inclusive jets are reconstructed by charged hadrons, while the D$^0$-tagged jets are reconstructed by charged hadrons and neutral D$^0$ mesons following the ALICE measurement~\cite{Dhankher:2024rkv}. Charged hadrons in the jets are required to have $|\eta| < 0.9$. The anti-$k_{\rm T}$ algorithm implemented in the FastJet package~\cite{Cacciari:2011ma} is used with jet radius parameter $R = 0.4$. The reconstructed D$^0$-tagged jets must contain at least one D$^0$ meson with $\pTD > 5$ GeV/c. Inclusive jets are selected by requiring the leading particle to satisfy $\pT>5.33$ GeV/c, the same cut on the transverse mass $m_{\rm T} = \sqrt{p_{\rm T}^2+m_0^2}$ as D$^0$-tagged jets, where $m_0$ is the particle mass and assumed to be 0 for leading particles in inclusive jets.
In \fig{fig:D0_baseline}, we show the angularity distributions of D$^0$-tagged and inclusive jets for four different values of $\alpha$: 1.0, 1.5, 2.0, and 3.0 in p+p collisions at $\sqrt{s}=5.02$ TeV compared with the ALICE preliminary data~\cite{Dhankher:2024rkv}. Reasonably good agreements are seen and one can also observe that the D$^0$-tagged (charm enriched) jets have distinctly narrower angularity distributions than inclusive jets for small $\alpha$, where the mass effects can be critical. From the ratios of D$^0$-tagged to inclusive jets shown in the bottom panels, one can see that our calculations provide a good description of the ALICE data for all values of $\lambda^{\kappa=1}_\alpha$, except for a slight overestimation of the ratios near $\lambda^{\kappa=1}_\alpha\sim$ 0. As the value of $\alpha$ increases, shapes of $\lambda^{\kappa=1}_\alpha$ for D$^0$-tagged and inclusive jets begin to converge since wide-angle emissions get more weights with larger $\alpha$ values.

\bef
\hspace{-0.1in}
  \includegraphics[width=1\linewidth]{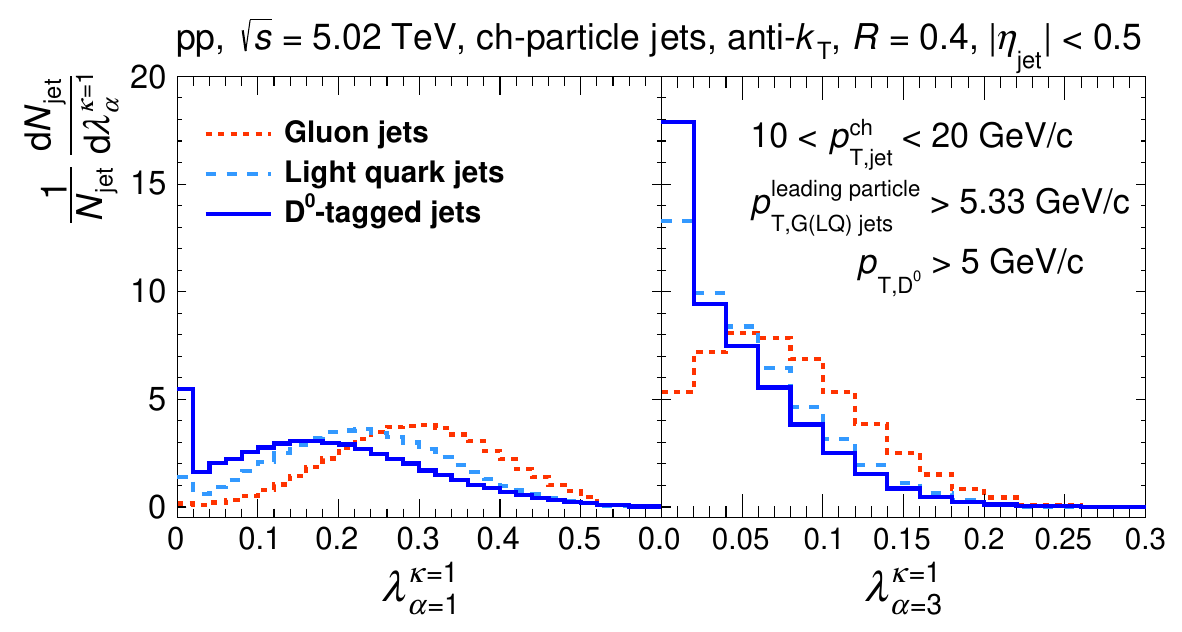}
  \caption{ (Color online) Normalized angularity distributions of light-quark and gluon jets for $\alpha=$ 1 and $\alpha=$ 3 with the fixed $\kappa=$ 1.0. The angularity distributions of D$^0$-tagged jets are also shown for comparison.}
  \label{fig:pt1_R04_class}
\eef

To explore the origin of different angularities for inclusive and D$^0$-tagged jets, we investigate the relation between the jet angularity and its flavor. In this work, we employ an infrared-and-collinear-safe jet flavor algorithm, the Interleaved Flavor Neutralization (IFN)~\cite{Caola:2023wpj}, to identify the jet flavor. Although such a flavor identification is not accessible experimentally, it is still informative to carry out such studies in Monte Carlo simulations. Note that for the production of heavy-quark jets, the pair creation process dominates the low-$\pT$ region (above $90\%$ at $10 < \pTJ < 20$ GeV/c), while the gluon splitting will be important at higher $\pT$ ($p_{\rm T,jet}>$100 GeV/c) \cite{Banfi:2007gu, Wang:2024yag}. Additionally, for inclusive jets, they are dominated by light-quark and gluon jets. In \fig{fig:pt1_R04_class}, we show the angularity distributions of these three types of jets in p+p collisions for $\alpha = 1.0$ (left) and $\alpha = 3.0$ (right). The peaks near $\lambda^{\kappa}_{\alpha}\sim$ 0 denote jets with only one component. We find that gluon jets have the broadest angularity distributions compared to the light-quark and D$^0$-tagged jets. This is because the larger Casimir color factor for gluons leads to more vacuum radiations than quarks. We also observe that the D$^0$-tagged jets have a narrower angularity distribution compared to the light-quark jets; such a difference is likely related to the ``dead-cone'' effect \cite{Dokshitzer:2001zm, ALICE:2019nuy}, which suppresses small angle radiation for heavy quarks. As $\alpha$ increases, the differences in angularities between the D$^0$-tagged and light-quark jets tend to disappear, but the gluon and light-quark jet angularities are still well separated.

\section{Jet quenching framework}
\label{framework}
To simulate the in-medium evolution of both inclusive and heavy-flavor jets, we utilize the
Simulating Heavy quark Energy Loss with Langevin equations (SHELL) model~\cite{Dai:2018mhw, Wang:2019xey, Wang:2020qwe, Wang:2020ukj, Li:2024uzk}, which can simultaneously take into account the elastic and inelastic partonic scatterings in the QGP medium. We use the p+p event with a complete vacuum parton shower generated by the POWHEG+PYTHIA8 event generator as the input to the SHELL model. The initial spatial distribution of the partons is sampled using a Monte Carlo Glauber model~\cite{Miller:2007ri}. Since the elastic scatterings between heavy quarks and thermal partons can be treated as Brownian motion, the transport of heavy quarks in the hot and dense expanding medium can be well described by the modified Langevin equations as follows

\begin{eqnarray}
&&    \vec{x}(t + \Delta t) = \vec{x}(t) + \frac{\vec{p}(t)}{E}\Delta t, \\
&&    \vec{p}(t + \Delta t) = \vec{p}(t) - \eta_D\vec{p}(t)\Delta t+\vec{\xi}(t)\Delta t-\vec{p}_{\rm g},
\label{eq:lang2}
\end{eqnarray}
where $\Delta t$ is the time step of the Monte Carlo simulation, $\vec{x}(t)$ and $\vec{p}(t)$ are the position and momentum of partons at time $t$, respectively. $E$ is the energy of the parton, and $\eta_D$ is the drag coefficient, which controls the energy-dissipation strength of heavy quarks in the medium. The stochastic term $\xi(t)$ represents the random kicks on heavy quarks from the thermal partons, which obeys the Gaussian distribution $\langle \xi^{i}(t) \xi^{j}(t^{\prime}) \rangle = \kappa \delta^{ij} \delta(t - t^{\prime})$ with $\kappa$ being the momentum diffusion coefficient that can be related to $\eta_D$ using the fluctuation-dissipation theorem $\kappa=2\eta_D ET$, where $T$ is the temperature of the medium. In the simulations, we set $2\pi T D_s = 4.0$, where $D_s$ is the spatial diffusion coefficient, which is extracted by a $\chi^2$ fitting to the yield suppression of D mesons measured by CMS \cite{CMS:2017qjw} and ALICE \cite{ALICE:2018lyv}.
On the right-hand side of \eq{eq:lang2}, the last term $-p_g$ denotes momentum shift caused by medium-induced gluon radiation. In our framework, we employ the higher-twist \cite{Guo:2000nz, Zhang:2003wk, Zhang:2003yn, Majumder:2009ge} formalism to simulate the medium-induced gluon radiation, whose
spectrum is given as follows:

\begin{eqnarray}
\frac{dN_g}{ dxdk^{2}_{\perp}dt}=\frac{2\alpha_{s} C_s P(x)\hat{q}}{\pi k^{4}_{\perp}}\sin^2(\frac{t-t_i}{2\tau_f})(\frac{k^2_{\perp}}{k^2_{\perp}+x^2M^2})^4,
\label{eq:dndxk}
\end{eqnarray}
where $x$ and $k_{\perp}$ are the gluon's energy fraction with respect to the parent parton and its transverse momentum, respectively. $\alpha_{s}$ is the strong coupling constant, $C_s$ is the quadratic Casimir factor in color representation, $P(x)$ is the QCD splitting function, $t_i$ represents the initial moment when the radiation begins, $\tau_f=2Ex(1-x)/(k^2_\perp+x^2M^2)$ is the formation time of the daughter gluon considering the Landau-Pomeranchuk-Migdal (LPM) effects \cite{Wang:1994fx,Zakharov:1996fv}, and $\hat{q} \propto q_0 (T/T_0)^3$ is the medium's transport coefficient \cite{Chen:2010te}. In the simulations, we set $q_0 = 1.2$~GeV$^2$/fm which is extracted based on the identified hadron productions in A+A collisions as done in our precious work~\cite{Ma:2018swx}. The last term in \eq{eq:dndxk} denotes the suppression factor resulting from the ``dead-cone'' effect, which reduces small angle ($\theta < M_Q/E$, $M_Q$ being the mass of heavy quarks) gluon radiation.

In the calculation, we utilize the CLVisc hydrodynamic model \cite{Pang:2016igs, Pang:2018zzo} to generate discrete temperature and velocity of the medium in a space-time grid of the expanding fireball. Then, the local temperature and velocity of the medium at which a parton passes through can be obtained by linear interpolation on its time and position according to this grid. When the local temperature reaches the presumptive critical temperature $T_c = 0.165$ GeV, the parton fragments into hadrons with the Colorless Hadronization prescription, which was developed by the JETSCAPE collaboration \cite{Putschke:2019yrg} based on the Lund string fragmentation model \cite{Andersson:1983jt}.

As the energetic parton propagates through the QGP, it loses energy, which is gained by the medium and excites a wake that is correlated with the direction of the parton \cite{Chesler:2007an}. After freeze-out, this wake becomes soft hadrons that are reconstructed as part of the jet, and it is challenging to isolate such medium response experimentally.
Since the jet angularities are closely related to the transverse momenta and opening angles of jet constituents, this medium response effect needs to be considered for a fair comparison to data. We follow the approach of implementing the medium response effect in the Hybrid Model \cite{Casalderrey-Solana:2016jvj}, which performs an expansion of the Cooper-Frye formula \cite{Cooper:1974mv} at
the perturbed freeze-out hypersurface and yields an estimate of the soft hadron spectrum from wake \cite{Casalderrey-Solana:2016jvj}:

\begin{align}
  \label{eq:wake}
  E\frac{d \Delta N}{d^3p}=&\frac{1}{32 \pi} \, \frac{m_{\rm T}}{T^5} \, \cosh(y-y_j) \exp\left[-\frac{m_{\rm T}}{T}\cosh(y-y_j)\right] \notag \\
      &\times \Big\{ \pT \Delta P_{\rm T} \cos (\phi-\phi_j) \notag \\
      &+\frac{1}{3}m_{\rm T} \, \Delta M_{\rm T} \, \cosh (y-y_j) \Big\},
\end{align}
where $m_{\rm T}$, $\pT$, $y$, and $\phi$ are the transverse mass, transverse momentum, rapidity, and azimuthal angle of the emitted wake particles, respectively. $y_j$ and $\phi_j$ are the rapidity and azimuthal angle of the energetic parton, respectively. $T$ denotes the freeze-out temperature of the hot QCD medium. $\Delta M_{\rm T} = \Delta E / y_j$ and $\Delta P_{\rm T}$ are the transverse mass and transverse momentum transferred from the jet to the medium, where $\Delta E$ is the jet energy loss. The SHELL model has been successfully employed in the studies of heavy-flavor jet production in high-energy nuclear collisions, such as the yield suppression~\cite{Wang:2023udp, ATLAS:2022agz}, transverse momentum balance~\cite{Dai:2018mhw, Li:2024uzk}, angular correlations~\cite{Wang:2020qwe, Wang:2021jgm} and substructure modification~\cite{Wang:2019xey, Wang:2020ukj, Li:2022tcr, CMS:2019jis}

\bef
  \includegraphics[width=0.85\linewidth]{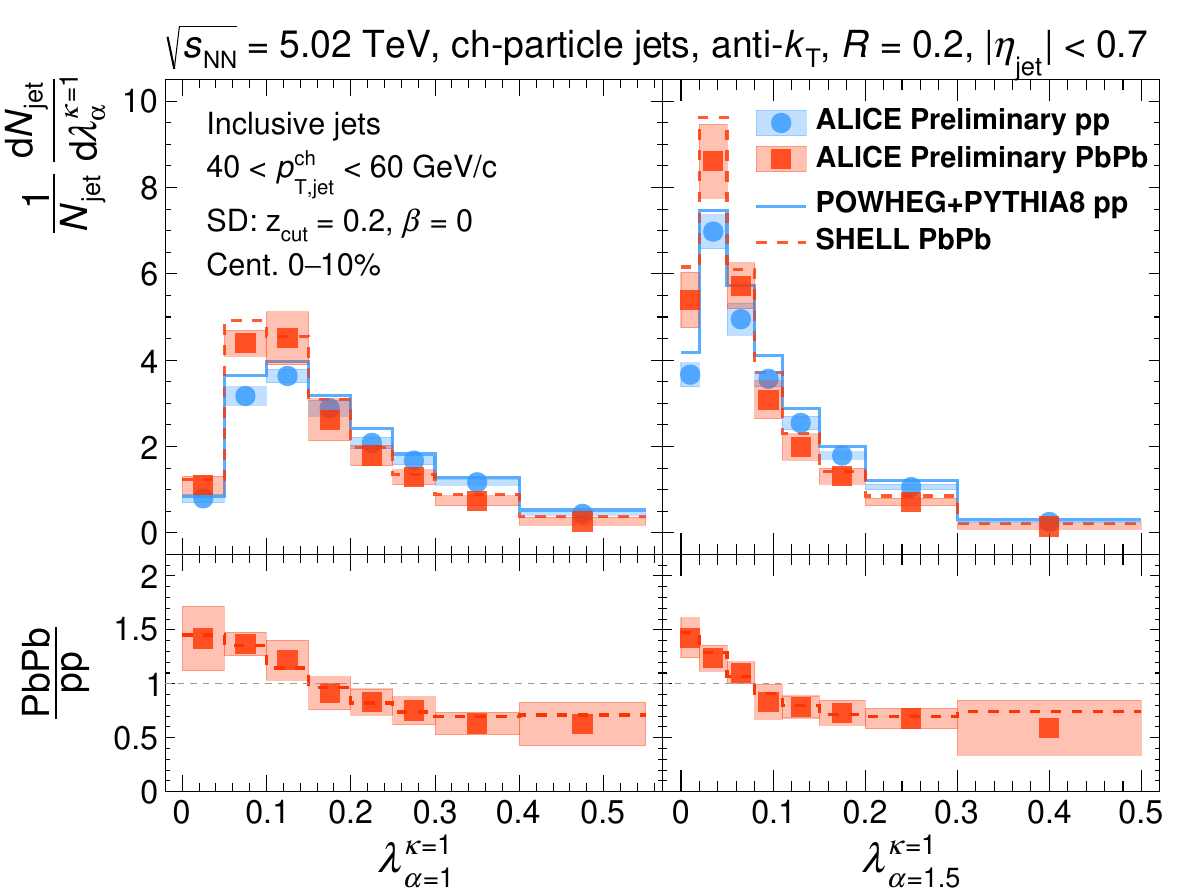}
  \includegraphics[width=0.85\linewidth]{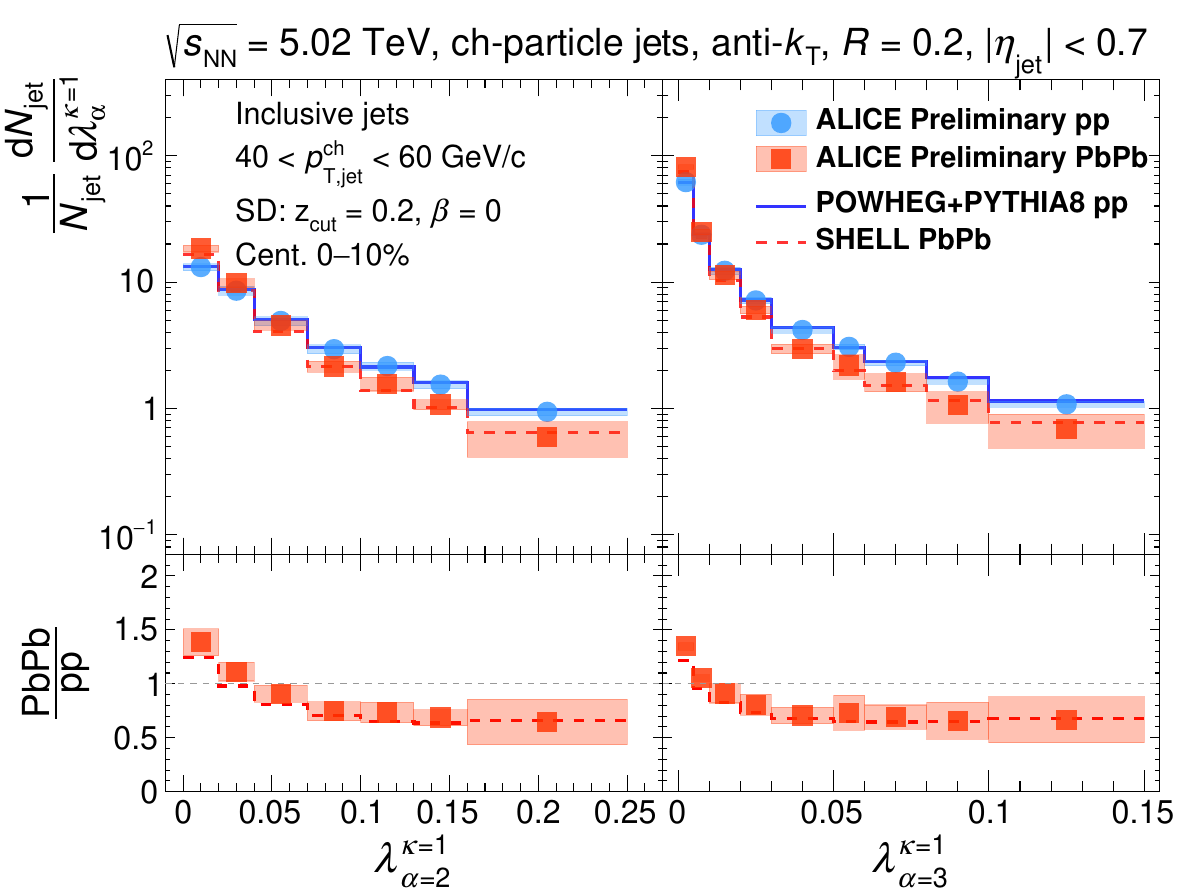}
  \caption{ (Color online) Normalized angularity distributions of inclusive jets for four different values of $\alpha$: 1.0, 1.5, 2.0, and 3.0 with the fixed $\kappa=$ 1.0 in p+p and 0-10\% Pb+Pb collisions at $\sqrt{s_{\rm NN}} = 5.02$ TeV, compared to the ALICE preliminary data~\cite{Vertesi:2024tdv}. The ratios of PbPb/pp are also shown in the lower panels.}
  \label{fig:RAA_baseline}
\eef

\section{Numerical Results and Discussions}
\label{results}

To validate our model as introduced in the last section, we calculate the angularities of inclusive jets in p+p and 0-10\% Pb+Pb collisions at $\sqrt{s_{\rm NN}} = 5.02$ TeV, and compare to the ALICE preliminary data~\cite{Vertesi:2024tdv} as shown in \fig{fig:RAA_baseline}. In line with the ALICE measurement, inclusive jets are reconstructed using charged particles with anti-$k_{\rm T}$ algorithm for $R = 0.2$ in the kinematic range $40 < \pTJ < 60$ GeV/c. Then, the inclusive jets are groomed by soft-drop \cite{Larkoski:2014wba} with parameters $\beta = 0$ and $z_{\rm cut} = 0.2$. We find that the $\lambda^{\kappa=1}_{\alpha}$ distributions calculated by POWHEG+PYTHIA8 (p+p) and SHELL model (Pb+Pb) both give decent descriptions to the ALICE data for all $\alpha$. From the ratio of Pb+Pb over pp, we observe enhancement at small $\lambda^{\kappa=1}_{\alpha}$ and suppression at large $\lambda^{\kappa=1}_{\alpha}$ in Pb+Pb collisions relative to p+p, consistent with the previous study \cite{Yan:2020zrz}. The modification pattern of $\lambda^{\kappa=1}_{\alpha}$ distribution in Pb+Pb collisions can be explained by the ``selection bias'' effect \cite{Renk:2012ve, Brewer:2021hmh, Wang:2021jgm, Wang:2024plm, Kang:2023ycg}. With the steeply falling jet spectrum, the effectively quenched jets may have a lower probability of passing the $\pT$ selection threshold in A+A collisions due to the energy loss in the QGP, while the one with insufficient quenching survives, referred to as the ``selection bias''. Such bias might enhance the fraction of the initially narrow jets in the selected sample in A+A collisions, different from the sample selected in p+p. As shown in \fig{fig:pt1_R04_class}, quark jets have a narrower angularity distribution than gluon jets in vacuum. Since gluon jets suffer more energy loss than quark jets in nuclear-nuclear collisions, quark jets have a larger probability of surviving the kinematic selection in Pb+Pb collisions. Therefore, an increased fraction of survived narrower quark jets can lead to an enhancement at small $\lambda^{\kappa=1}_{\alpha}$ and a suppression at large $\lambda^{\kappa=1}_{\alpha}$ region in Pb+Pb collisions relative to p+p. More detailed discussions about the influence of the selection bias on the jet substructure modifications in heavy-ion collisions can be found in Refs. \cite{Wang:2024plm, Kang:2023ycg, Brewer:2021hmh, Renk:2012ve}.

\begin{figure*}[!t]
\centering
\includegraphics[width=0.55\linewidth]{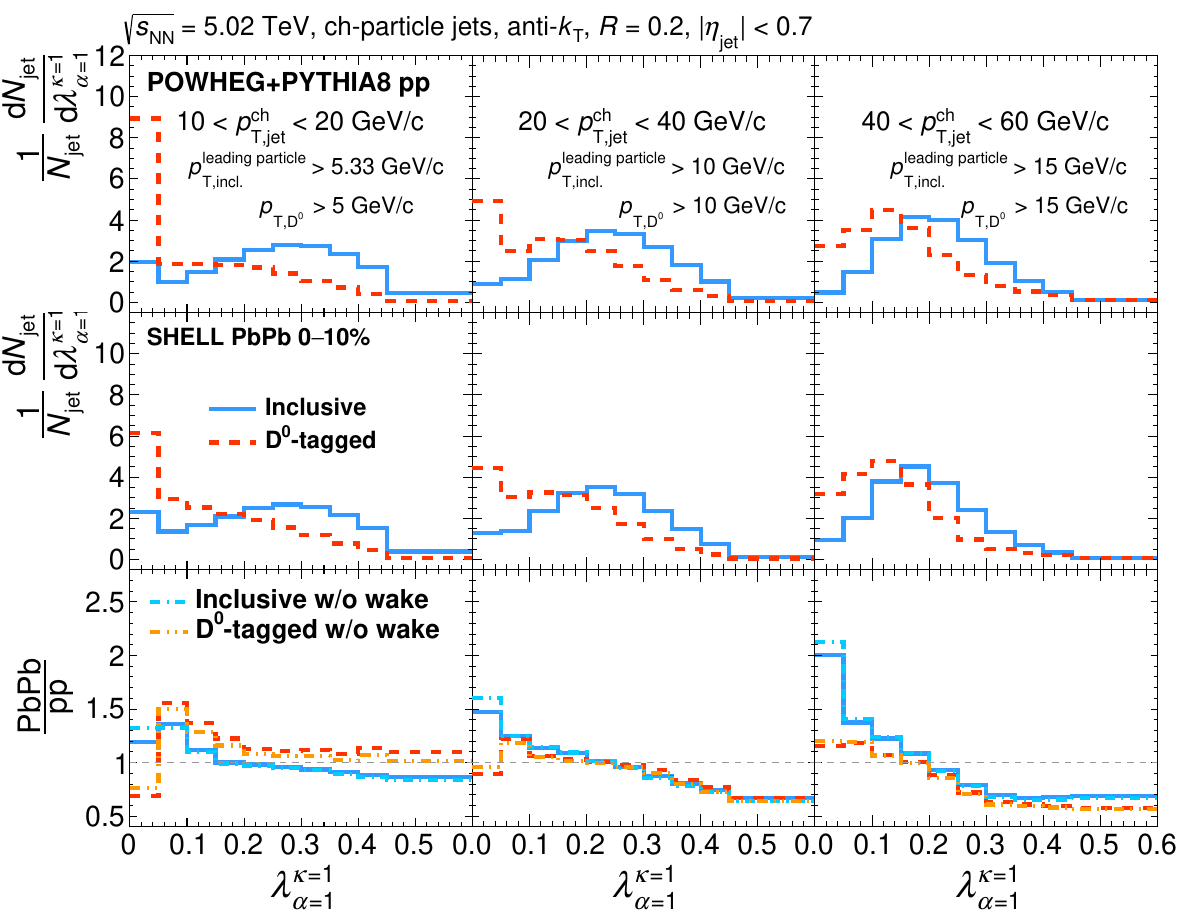}
\caption{ (Color online) Normalized angularity distributions of the D$^0$-tagged and inclusive jets for three $\pTJ$ intervals: $10 < \pTJ < 20$ GeV/c, $20 < \pTJ < 40$ GeV/c and $40 < \pTJ < 60$ GeV/c (from left to right panels) in p+p (top panels) and 0-10\% Pb+Pb (middle panels) collisions at $\sqrt{s_{\rm NN}} = 5.02$ TeV, as well as the ratios of Pb+Pb (with or without wake) over p+p (bottom panels).}
\label{fig:kappa1_R02}
\end{figure*}

Next, we turn to medium modifications of heavy-flavor jet angularities in nucleus-nucleus collisions at the LHC, which have not been measured to date. In \fig{fig:kappa1_R02}, we show the calculations of the jet angularities of inclusive and D$^0$-tagged jets in p+p and 0-10\% Pb+Pb collisions at $\sqrt{s_{\rm NN}} = 5.02$ TeV for three $\pTJ$ intervals: $10 < \pTJ < 20$ GeV/c, $20 < \pTJ < 40$ GeV/c and $40 < \pTJ < 60$ GeV/c, as well as the ratios of PbPb/pp. Note that in the latter two $\pTJ$ intervals the leading hadron (D$^0$ meson) in inclusive (D$^0$-tagged) jets is required to have $\pT>10$ GeV/c, and $\pT>15$ GeV/c, respectively. For $\pTJ > 20$ GeV/c, jet angularities get narrower in Pb+Pb collisions compared to those in p+p collisions for both D$^0$-tagged and inclusive jets, caused mainly by the selection bias. At $10 < \pTJ < 20$ GeV/c, we observe that the angularity distributions of D$^0$-tagged jets get slightly broader in Pb+Pb collisions compared to those in p+p collisions due to the jet quenching effect. Here, the lower $\pT$ threshold of jet selection can suppress the influence of the selection bias. Since the in-medium radiation of charm quarks reduces the fraction of jets containing only D$^0$ mesons, a remarkable suppression of D$^0$-tagged jet angularity near $\lambda^{\kappa=1}_{\alpha=1}\sim$ 0 can be found. Modifications of $\lambda_{\alpha=1}^{\kappa=1}$ distributions for inclusive jets are more distinct than those for the D$^0$-tagged jets at $\pTJ >$ 20 GeV/c. It can be explained by the fact that inclusive jets contain a considerable fraction of gluon jets, which is subject to the selection bias to a greater extent compared to the quark jet~\cite{Wang:2024plm, Kang:2023ycg}. The effect of medium response can be studied by comparing angularity distributions with and without including wake particles sampled with Eq. (\ref{eq:wake}). We find limited impact of medium response on jet angularities for jet radius of $R=0.2$. In the following discussions, we will focus on the medium modification of jet angularities at lower jet $\pT$ ($10 < \pTJ < 20$ GeV/c) where the jet quenching effect is pronounced and the influence of selection bias is moderate.

\bef
\includegraphics[width=0.85\linewidth]{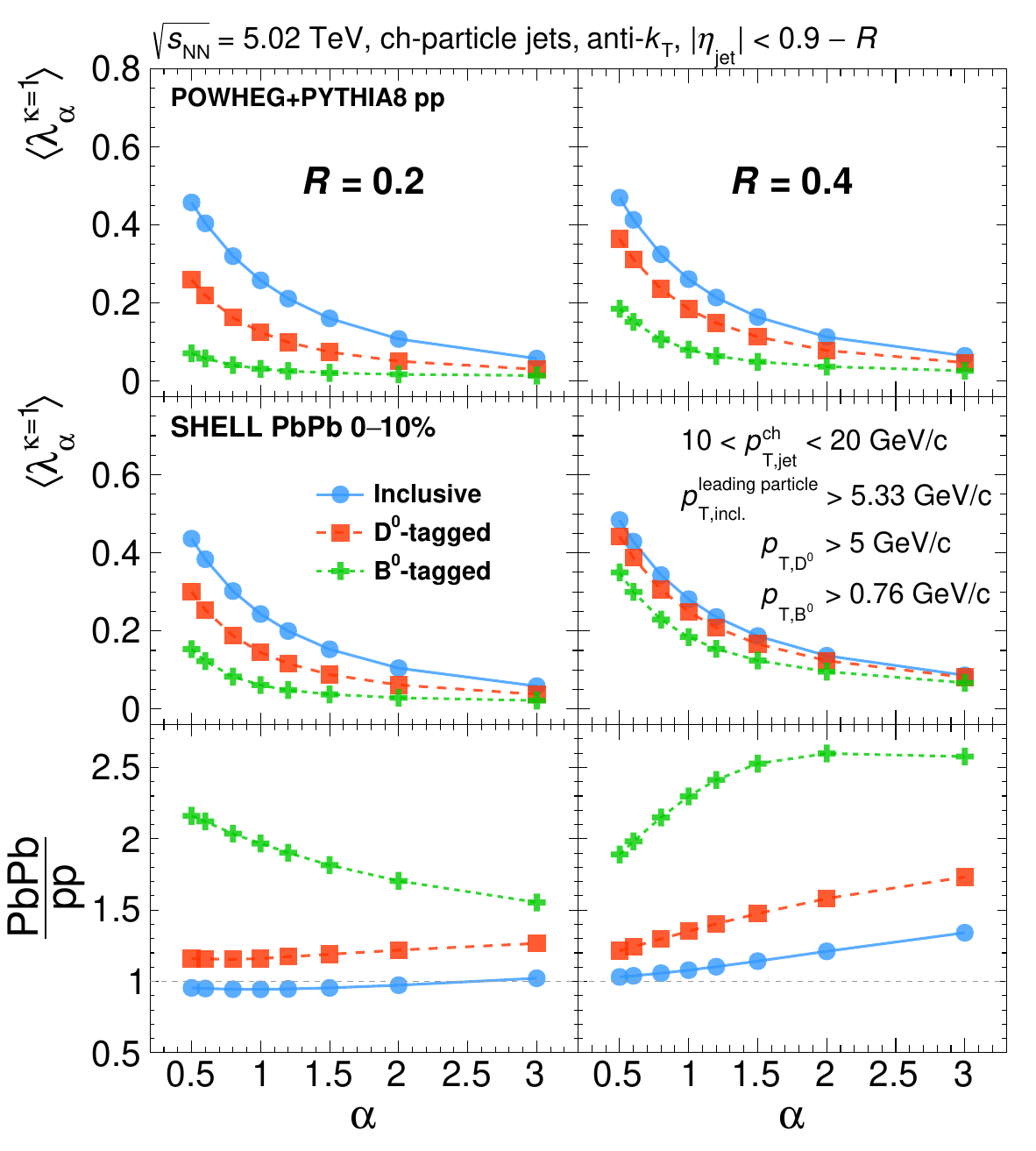}
\caption{ (Color online) Average angularities $\langle \lambda^{\kappa=1}_{\alpha} \rangle$ of inclusive, D$^0$-tagged and B$^0$-tagged jets within $10 < \pTJ < 20$ GeV/c for two jet radii $R = 0.2$ (left panels) and $R = 0.4$ (right panels) in p+p (top panels) and 0-10\% Pb+Pb (middle panels) collisions as a function of $\alpha$, as well as the ratios of Pb+Pb over p+p (bottom panels). }
\label{fig:alpha_avg}
\eef

\bef
\includegraphics[width=0.85\linewidth]{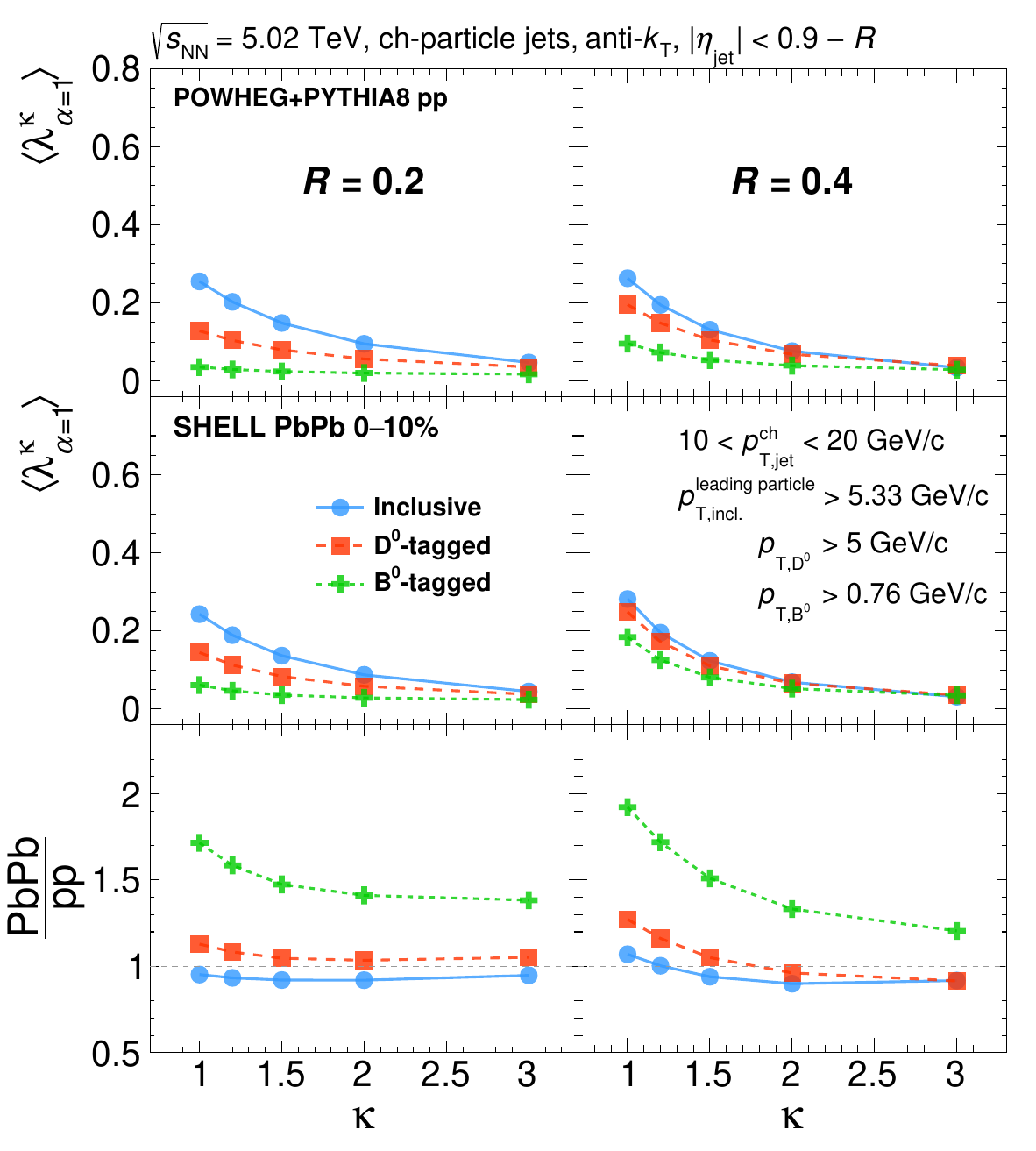}
\caption{ (Color online) Average angularities $\langle \lambda^{\kappa=1}_{\alpha} \rangle$ of inclusive, D$^0$-tagged and B$^0$-tagged jets within $10 < \pTJ < 20$ GeV/c for two jet radii $R = 0.2$ (left panels) and $R = 0.4$ (right panels) in p+p (top panels) and 0-10\% Pb+Pb (middle panels) collisions as a function of $\kappa$, as well as the ratios of Pb+Pb over p+p (bottom panels). }
\label{fig:kappa_avg}
\eef

Since the angularity values depend on both the momentum fraction and opening angle of particles in jets, it is of great interest to explore the mass dependence of angularity modification with respect to the two corresponding parameters $\kappa$ and $\alpha$ for inclusive and heavy-flavor jets. Hence, we also include calculations of average angularities for B$^0$-tagged jets that mainly originate from bottom quarks. In \fig{fig:alpha_avg}, we show average angularity values $\langle \lambda^{\kappa}_{\alpha} \rangle$ of inclusive, D$^0$-tagged and B$^0$-tagged jets as a function of the parameter $\alpha$ for $R = 0.2$ and $R = 0.4$ in p+p and 0-10\% Pb+Pb collisions at $\sqrt{s_{\rm NN}}=$ 5.02 TeV, as well as the ratios of PbPb/pp. In p+p collisions, $\langle \lambda^{\kappa}_{\alpha} \rangle$ decreases with the quark mass. This is because heavier quarks suffer from stronger influence of the ``dead-cone'' effect with less gluon radiation during parton shower leading to smaller angularities. In Pb+Pb collisions, the ordering of $\langle \lambda^{\kappa}_{\alpha} \rangle$ values for inclusive, D$^0$-tagged and B$^0$-tagged jets is the same as that in pp. However, we observe that the ratios of PbPb/pp have a reversed ordering, {\it i.e.}, B$^0$-tagged jets have the largest ratio while inclusive jets have the smallest one. Because B$^0$-tagged jets have smaller initial $\langle \lambda^{\kappa}_{\alpha} \rangle$ values, they are more sensitive to jet-medium interactions, such as $\pT$ broadening, medium-induced gluon radiation, which lead to broader jet angularities. For the larger jet radius $R = 0.4$ shown in the right panels, since more wide-angle particles from medium-induced radiation and the medium response are included, one can see that the angularity broadening of heavy-flavor jets get more significant compared to $R = 0.2$. In \fig{fig:kappa_avg}, we also test the sensitivity of jet angularity modifications to the parameter $\kappa$. Similarly, we obtain the same order of modification strength of angularities (PbPb/pp) for the inclusive and heavy-flavor jets in Pb+Pb collisions. However, since $\kappa$ controls the weight of particle momentum fraction, higher $\kappa$ values will suppress the contribution of soft particles in jets. Therefore, the ratios of PbPb/pp decrease with $\kappa$ for all types of jets.

\section{Summary}

We present the first theoretical study of heavy-flavor jet angularities in Pb+Pb collisions at $\sqrt{s_{\rm NN}}=$ 5.02 TeV. We use POWHEG+PYTHIA8 to simulate initial
heavy-flavor jet production in p+p collisions and employ the SHELL transport model to simulate the jet-medium interactions in nucleus-nucleus collisions. In p+p collisions, our calculated angularity distributions of the D$^0$-tagged jets are narrower than those of inclusive jets, consistent with the ALICE preliminary data. This is because inclusive jets contain a considerable fraction of gluon jets, which have broader angularity distributions than quark jets, while the D$^0$-tagged jets are mainly initiated by charm quarks. Additionally, the ``dead-cone'' effect suppresses gluon radiation at small angles for heavy quarks, resulting in a narrower angularity distribution compared to light-quark jets. In Pb+Pb collisions at $\sqrt{s_{\rm NN}}=$ 5.02 TeV, our calculations show that jet-medium interactions in the QGP widen the angularity distributions of $R = 0.2$ D$^0$-tagged jets compared to the p+p baseline at $10<\pTJ<20$ GeV/c. Furthermore, we notice that the angularity distributions of D$^0$-tagged and inclusive jets for $R = 0.2$ get narrower in Pb+Pb collisions compared to p+p collisions at $\pTJ>20$ GeV/c where the influence of the selection bias becomes significant, consistent with previous measurements at the LHC. By comparing average angularity values, $\langle \lambda^{\kappa}_{\alpha} \rangle$, of the inclusive, D$^0$-tagged and B$^0$-tagged jets with varying $\alpha$ and $\kappa$, one can see that the larger the quark mass is, the narrower the jet's angularities are. As a result of the narrower initial distribution, we predict that the heavy-flavor jets, especially the B$^0$-tagged ones, will experience stronger modifications of $\langle \lambda^{\kappa}_{\alpha} \rangle$ in Pb+Pb relative to the p+p baseline as compared to inclusive jets. In addition, a more significant broadening of jet angularities could be observed for larger jet radius because of enhanced contribution of wide-angle particles. It will be informative to test the predicted angularity broadening of the heavy-flavor jets at low $\pT$ and for large radius with improved detector capability and increased luminosity at the LHC.

\textbf{Acknowledgments:}
This research is supported by the Guangdong Major Project of Basic and Applied Basic Research No. 2020B0301030008, and the National Natural Science Foundation of China with Project Nos.~11935007 and 12035007. S. W. is further supported by the Open Foundation of Key Laboratory of Quark and Lepton Physics (MOE) No. QLPL2023P01 and the Talent Scientific Star-up Foundation of the China Three Gorges University (CTGU) with No. 2024RCKJ013.


\begin{thebibliography}{99}

\bibitem{Freedman:1976ub}
  B.~A.~Freedman and L.~D.~McLerran,
  Phys. Rev. D \textbf{16} (1977), 1169

\bibitem{Shuryak:1977ut}
  E.~V.~Shuryak,
  Sov. Phys. JETP \textbf{47} (1978), 212-219
  IYF-77-34.

\bibitem{Wang:1992qdg}
X.~N.~Wang and M.~Gyulassy,
Phys. Rev. Lett. \textbf{68} (1992), 1480-1483

\bibitem{Gyulassy:2003mc}
M.~Gyulassy, I.~Vitev, X.~N.~Wang and B.~W.~Zhang,
[arXiv:nucl-th/0302077 [nucl-th]].

\bibitem{Wang:2002ri}
E.~Wang and X.~N.~Wang,
Phys. Rev. Lett. \textbf{89}, 162301 (2002)
[arXiv:hep-ph/0202105 [hep-ph]].

\bibitem{Vitev:2009rd}
  I.~Vitev and B.~W.~Zhang,
  Phys.\ Rev.\ Lett.\  {\bf 104}, 132001 (2010).

\bibitem{Neufeld:2010fj}
  R.~B.~Neufeld, I.~Vitev and B.-W.~Zhang,
  Phys.\ Rev.\ C {\bf 83}, 034902 (2011).

\bibitem{Vitev:2008rz}
  I.~Vitev, S.~Wicks and B.~W.~Zhang,
  JHEP {\bf 0811}, 093 (2008).

\bibitem{He:2020iow}
Y.~He, L.~G.~Pang and X.~N.~Wang,
Phys. Rev. Lett. \textbf{125}, no.12, 122301 (2020)
[arXiv:2001.08273 [hep-ph]].

\bibitem{Chen:2022kic}
S.~Y.~Chen, J.~Yan, W.~Dai, B.~W.~Zhang and E.~Wang,
Chin. Phys. C \textbf{46} (2022) no.10, 104102
[arXiv:2204.01211 [hep-ph]].

\bibitem{Zhao:2021vmu}
W.~Zhao, W.~Ke, W.~Chen, T.~Luo and X.~N.~Wang,
Phys. Rev. Lett. \textbf{128} (2022) no.2, 022302
[arXiv:2103.14657 [hep-ph]].

\bibitem{Yang:2022nei}
Z.~Yang, T.~Luo, W.~Chen, L.~G.~Pang and X.~N.~Wang,
Phys. Rev. Lett. \textbf{130} (2023) no.5, 052301
[arXiv:2203.03683 [hep-ph]].

\bibitem{Yang:2023dwc}
Z.~Yang, Y.~He, I.~Moult and X.~N.~Wang,
Phys. Rev. Lett. \textbf{132} (2024) no.1, 1
[arXiv:2310.01500 [hep-ph]].

\bibitem{Zhang:2021xib}
H.~X.~Zhang, Y.~X.~Xiao, J.~W.~Kang and B.~W.~Zhang,
Nucl. Sci. Tech. \textbf{33}, no.11, 150 (2022)
[arXiv:2102.11792 [hep-ph]].

\bibitem{Xie:2024xbn}
M.~Xie, Q.~F.~Han, E.~K.~Wang, B.~W.~Zhang and H.~Z.~Zhang,
Nucl. Sci. Tech. \textbf{35}, no.7, 125 (2024)

\bibitem{JETSCAPE:2022jer}
A.~Kumar \textit{et al.} [JETSCAPE],
Phys. Rev. C \textbf{107} (2023) no.3, 034911
[arXiv:2204.01163 [hep-ph]].

\bibitem{Luo:2023nsi}
T.~Luo, Y.~He, S.~Cao and X.~N.~Wang,
Phys. Rev. C \textbf{109} (2024) no.3, 034919
[arXiv:2306.13742 [nucl-th]].

\bibitem{Zhang:2023oid}
S.~L.~Zhang, E.~Wang, H.~Xing and B.~W.~Zhang,
Phys. Lett. B \textbf{850}, 138549 (2024)
[arXiv:2303.14881 [hep-ph]].

\bibitem{Chen:2024cgx}
S.~Y.~Chen, K.~M.~Shen, W.~Dai, B.~W.~Zhang and E.~K.~Wang,
[arXiv:2409.13996 [nucl-th]].

\bibitem{Xu:2014tda}
J.~Xu, J.~Liao and M.~Gyulassy,
Chin. Phys. Lett. \textbf{32} (2015) no.9, 092501
[arXiv:1411.3673 [hep-ph]].

\bibitem{Ma:2023zfj}
Y.~G.~Ma, L.~G.~Pang, R.~Wang and K.~Zhou,
Chin. Phys. Lett. \textbf{40} (2023) no.12, 122101
[arXiv:2311.07274 [nucl-th]].

\bibitem{Wu:2022vbu}
J.~Wu, S.~Cao and F.~Li,
Chin. Phys. Lett. \textbf{41} (2024) no.3, 031202
[arXiv:2208.14297 [nucl-th]].

\bibitem{Yan:2020zrz}
J.~Yan, S.~Y.~Chen, W.~Dai, B.~W.~Zhang and E.~Wang,
Chin. Phys. C \textbf{45}, no.2, 024102 (2021)
[arXiv:2005.01093 [hep-ph]].

\bibitem{Dong:2019unq}
  X.~Dong and V.~Greco,
  Prog.\ Part.\ Nucl.\ Phys.\  {\bf 104}, 97 (2019).

\bibitem{Tang:2020ame}
Z.~Tang, Z.~B.~Tang, W.~Zha, W.~M.~Zha, Y.~Zhang and Y.~F.~Zhang,
Nucl. Sci. Tech. \textbf{31} (2020) no.8, 81
[arXiv:2105.11656 [nucl-ex]].


\bibitem{Wang:2023eer}
S.~Wang, W.~Dai, E.~Wang, X.~N.~Wang and B.~W.~Zhang,
Symmetry \textbf{15} (2023), 727
[arXiv:2303.14660 [nucl-th]].

\bibitem{Zhao:2020jqu}
  J.~Zhao, K.~Zhou, S.~Chen and P.~Zhuang,
  Prog.\ Part.\ Nucl.\ Phys.\  {\bf 114} (2020) 103801
  [arXiv:2005.08277 [nucl-th]].

\bibitem{Francis:2015daa}
A.~Francis, O.~Kaczmarek, M.~Laine, T.~Neuhaus and H.~Ohno,
Phys. Rev. D \textbf{92} (2015) no.11, 116003
[arXiv:1508.04543 [hep-lat]].

\bibitem{Kumar:2020wvb}
A.~Kumar, A.~Majumder and J.~H.~Weber,
Phys. Rev. D \textbf{106} (2022) no.3, 034505
[arXiv:2010.14463 [hep-lat]].

\bibitem{JETSCAPE:2022hcb}
W.~Fan \textit{et al.} [JETSCAPE],
Phys. Rev. C \textbf{107}, no.5, 054901 (2023)
[arXiv:2208.00983 [nucl-th]].

\bibitem{Xu:2017obm}
Y.~Xu, J.~E.~Bernhard, S.~A.~Bass, M.~Nahrgang and S.~Cao,
Phys. Rev. C \textbf{97} (2018) no.1, 014907
[arXiv:1710.00807 [nucl-th]].

\bibitem{Li:2019lex}
S.~Li and J.~Liao,
Eur. Phys. J. C \textbf{80} (2020) no.7, 671
[arXiv:1912.08965 [hep-ph]].

\bibitem{Cao:2018ews}
S.~Cao {\it et al.},
Phys.\ Rev.\ C {\bf 99}, no. 5, 054907 (2019)
[arXiv:1809.07894 [nucl-th]].

\bibitem{CMS:2017qjw}
A.~M.~Sirunyan \textit{et al.} [CMS],
Phys. Lett. B \textbf{782} (2018), 474-496
[arXiv:1708.04962 [nucl-ex]].

\bibitem{PHENIX:2011img}
A.~Adare \textit{et al.} [PHENIX],
Phys. Rev. C \textbf{84}, 054912 (2011)
[arXiv:1103.6269 [nucl-ex]].

\bibitem{STAR:2013eve}
L.~Adamczyk \textit{et al.} [STAR],
Phys. Rev. C \textbf{90}, no.2, 024906 (2014)
[arXiv:1310.3563 [nucl-ex]].

\bibitem{ALICE:2014wnc}
B.~B.~Abelev \textit{et al.} [ALICE],
Phys. Lett. B \textbf{738}, 361-372 (2014)
[arXiv:1405.4493 [nucl-ex]].

\bibitem{Adamczyk:2017xur}
  L.~Adamczyk {\it et al.} [STAR Collaboration],
  Phys.\ Rev.\ Lett.\  {\bf 118}, no. 21, 212301 (2017)

\bibitem{Sirunyan:2017plt}
  A.~M.~Sirunyan {\it et al.} [CMS Collaboration],
  Phys.\ Rev.\ Lett.\  {\bf 120}, no. 20, 202301 (2018)

\bibitem{STAR:2019clv}
J.~Adam \textit{et al.} [STAR],
Phys. Rev. Lett. \textbf{123}, no.16, 162301 (2019)
[arXiv:1905.02052 [nucl-ex]].

\bibitem{STAR:2019ank}
J.~Adam \textit{et al.} [STAR],
Phys. Rev. Lett. \textbf{124} (2020) no.17, 172301
[arXiv:1910.14628 [nucl-ex]].

\bibitem{Vermunt:2019ecg}
L.~Vermunt [ALICE],
PoS \textbf{EPS-HEP2019} (2020), 297
[arXiv:1910.11738 [nucl-ex]].

\bibitem{Cacciari:2005rk}
M.~Cacciari, P.~Nason and R.~Vogt,
Phys. Rev. Lett. \textbf{95} (2005), 122001
[arXiv:hep-ph/0502203 [hep-ph]].

\bibitem{Kniehl:2004fy}
B.~A.~Kniehl, G.~Kramer, I.~Schienbein and H.~Spiesberger,
Phys. Rev. D \textbf{71} (2005), 014018
[arXiv:hep-ph/0410289 [hep-ph]].

\bibitem{Eskola:2009uj}
K.~J.~Eskola, H.~Paukkunen and C.~A.~Salgado,
JHEP \textbf{04} (2009), 065
[arXiv:0902.4154 [hep-ph]].

\bibitem{Eskola:2016oht}
K.~J.~Eskola, P.~Paakkinen, H.~Paukkunen and C.~A.~Salgado,
Eur. Phys. J. C \textbf{77} (2017) no.3, 163
[arXiv:1612.05741 [hep-ph]].

\bibitem{Cao:2016gvr}
S.~Cao, T.~Luo, G.~Y.~Qin and X.~N.~Wang,
Phys. Rev. C \textbf{94} (2016) no.1, 014909
[arXiv:1605.06447 [nucl-th]].

\bibitem{Li:2020umn}
S.~Li, W.~Xiong and R.~Wan,
Eur. Phys. J. C \textbf{80} (2020) no.12, 1113
[arXiv:2012.02489 [hep-ph]].

\bibitem{Xing:2023ciw}
W.~J.~Xing, S.~Cao and G.~Y.~Qin,
Phys. Lett. B \textbf{850}, 138523 (2024)
[arXiv:2303.12485 [hep-ph]].

\bibitem{Cao:2019iqs}
  S.~Cao, K.~J.~Sun, S.~Q.~Li, S.~Y.~F.~Liu, W.~J.~Xing, G.~Y.~Qin and C.~M.~Ko,
  Phys.\ Lett.\ B {\bf 807}, 135561 (2020)
  [arXiv:1911.00456 [nucl-th]].

\bibitem{Plumari:2017ntm}
S.~Plumari, V.~Minissale, S.~K.~Das, G.~Coci and V.~Greco,
Eur. Phys. J. C \textbf{78}, no.4, 348 (2018)
[arXiv:1712.00730 [hep-ph]].

\bibitem{He:2019vgs}
M.~He and R.~Rapp,
Phys. Rev. Lett. \textbf{124}, no.4, 042301 (2020)
[arXiv:1905.09216 [nucl-th]].

\bibitem{Li:2018xuv}
H.~T.~Li and I.~Vitev,
JHEP \textbf{07}, 148 (2019)
[arXiv:1811.07905 [hep-ph]].

\bibitem{Kang:2018wrs}
Z.~B.~Kang, J.~Reiten, I.~Vitev and B.~Yoon,
Phys.\ Rev.\ D {\bf 99}, no. 3, 034006 (2019)
[arXiv:1810.10007 [hep-ph]].

\bibitem{Dai:2018mhw}
W.~Dai, S.~Wang, S.~L.~Zhang, B.~W.~Zhang and E.~Wang,
Chin.\ Phys.\ C {\bf 44} (2020) no.10, 104105
[arXiv:1806.06332 [nucl-th]].

\bibitem{Li:2024uzk}
Y.~Li, S.~Shen, S.~Wang and B.~W.~Zhang,
Nucl. Sci. Tech. \textbf{35}, no.7, 113 (2024)
[arXiv:2401.01706 [hep-ph]].

\bibitem{Wang:2020qwe}
S.~Wang, W.~Dai, B.~W.~Zhang and E.~Wang,
Chin. Phys. C \textbf{47}, no.5, 054102 (2023)
[arXiv:2005.07018 [hep-ph]].

\bibitem{Wang:2019xey}
S.~Wang, W.~Dai, B.~W.~Zhang and E.~Wang,
Eur.\ Phys.\ J.\ C {\bf 79} (2019) no.9,  789
[arXiv:1906.01499 [nucl-th]].

\bibitem{Wang:2020ukj}
S.~Wang, W.~Dai, B.~W.~Zhang and E.~Wang,
Chin. Phys. C \textbf{45}, no.6, 064105 (2021)
[arXiv:2012.13935 [nucl-th]].

\bibitem{Li:2022tcr}
Y.~Li, S.~Wang and B.~W.~Zhang,
Phys. Rev. C \textbf{108}, no.2, 2 (2023)
[arXiv:2209.00548 [hep-ph]].

\bibitem{Li:2017wwc}
H.~T.~Li and I.~Vitev,
Phys. Lett. B \textbf{793}, 259-264 (2019)
[arXiv:1801.00008 [hep-ph]].

\bibitem{Ringer:2019rfk}
F.~Ringer, B.~W.~Xiao and F.~Yuan,
Phys. Lett. B \textbf{808} (2020), 135634
[arXiv:1907.12541 [hep-ph]].

\bibitem{JETSCAPE:2023hqn}
Y.~Tachibana \textit{et al.} [JETSCAPE],
[arXiv:2301.02485 [hep-ph]].

\bibitem{Milhano:2017nzm}
G.~Milhano, U.~A.~Wiedemann and K.~C.~Zapp,
Phys. Lett. B \textbf{779} (2018), 409-413
[arXiv:1707.04142 [hep-ph]].

\bibitem{Caucal:2019uvr}
P.~Caucal, E.~Iancu and G.~Soyez,
JHEP \textbf{10} (2019), 273
[arXiv:1907.04866 [hep-ph]].

\bibitem{Wang:2022yrp}
L.~Wang, J.~W.~Kang, Q.~Zhang, S.~Shen, W.~Dai, B.~W.~Zhang and E.~Wang,
Chin. Phys. Lett. \textbf{40}, no.3, 032101 (2023)
[arXiv:2211.13674 [nucl-th]].

\bibitem{ALICE:2022phr}
S.~Acharya \textit{et al.} [ALICE],
Phys. Rev. Lett. \textbf{131}, no.19, 192301 (2023)
[arXiv:2208.04857 [nucl-ex]].

\bibitem{Larkoski:2014pca}
A.~J.~Larkoski, J.~Thaler and W.~J.~Waalewijn,
JHEP \textbf{11} (2014), 129
[arXiv:1408.3122 [hep-ph]].

\bibitem{Dhankher:2024rkv}
P.~Dhankher [ALICE],
PoS \textbf{HardProbes2023}, 140 (2024)

\bibitem{Caletti:2021oor}
S.~Caletti, O.~Fedkevych, S.~Marzani, D.~Reichelt, S.~Schumann, G.~Soyez and V.~Theeuwes,
JHEP \textbf{07}, 076 (2021)
[arXiv:2104.06920 [hep-ph]].

\bibitem{ALICE:2021njq}
S.~Acharya \textit{et al.} [ALICE],
JHEP \textbf{05}, 061 (2022)
[arXiv:2107.11303 [nucl-ex]].

\bibitem{Reichelt:2021svh}
D.~Reichelt, S.~Caletti, O.~Fedkevych, S.~Marzani, S.~Schumann and G.~Soyez,
JHEP \textbf{03}, 131 (2022)
[arXiv:2112.09545 [hep-ph]].

\bibitem{Budhraja:2023rgo}
A.~Budhraja, R.~Sharma and B.~Singh,
[arXiv:2305.10237 [hep-ph]].

\bibitem{Chien:2024uax}
Y.~T.~Chien, O.~Fedkevych, D.~Reichelt and S.~Schumann,
JHEP \textbf{07}, 230 (2024)
[arXiv:2404.04168 [hep-ph]].

\bibitem{Frixione:2007vw}
S.~Frixione, P.~Nason and C.~Oleari,
JHEP \textbf{11}, 070 (2007)
[arXiv:0709.2092 [hep-ph]].

\bibitem{Alioli:2010xd}
S.~Alioli, P.~Nason, C.~Oleari and E.~Re,
JHEP \textbf{06}, 043 (2010)
[arXiv:1002.2581 [hep-ph]].

\bibitem{Bierlich:2022pfr}
C.~Bierlich, S.~Chakraborty, N.~Desai, L.~Gellersen, I.~Helenius, P.~Ilten, L.~L\"onnblad, S.~Mrenna, S.~Prestel and C.~T.~Preuss, \textit{et al.}
[arXiv:2203.11601 [hep-ph]].

\bibitem{Alioli:2010xa}
S.~Alioli, K.~Hamilton, P.~Nason, C.~Oleari and E.~Re,
JHEP \textbf{04}, 081 (2011)
[arXiv:1012.3380 [hep-ph]].

\bibitem{Hou:2019qau}
T.~J.~Hou, K.~Xie, J.~Gao, S.~Dulat, M.~Guzzi, T.~J.~Hobbs, J.~Huston, P.~Nadolsky, J.~Pumplin and C.~Schmidt, \textit{et al.}
[arXiv:1908.11394 [hep-ph]].

\bibitem{Cacciari:2011ma}
M.~Cacciari, G.~P.~Salam and G.~Soyez,
Eur. Phys. J. C \textbf{72} (2012), 1896
[arXiv:1111.6097 [hep-ph]].

\bibitem{Caola:2023wpj}
F.~Caola, R.~Grabarczyk, M.L.~Hutt et al.,
Phys. Rev. D \textbf{108}, 094010 (2023) 
[arXiv:2306.07314 [hep-ph]].

\bibitem{Banfi:2007gu}
A.~Banfi, G.~P.~Salam and G.~Zanderighi,
JHEP \textbf{07} (2007), 026
[arXiv:0704.2999 [hep-ph]].

\bibitem{Wang:2024yag}
S.~Wang, S.~Li, Y.~Li, B.~W.~Zhang and E.~Wang,
[arXiv:2410.21834 [hep-ph]].

\bibitem{Dokshitzer:2001zm}
Y.~L.~Dokshitzer and D.~E.~Kharzeev,
Phys. Lett. B \textbf{519} (2001), 199-206
[arXiv:hep-ph/0106202 [hep-ph]].

\bibitem{ALICE:2019nuy}
S.~Acharya \textit{et al.} [ALICE],
Phys. Lett. B \textbf{804} (2020), 135377
[arXiv:1910.09110 [nucl-ex]].

\bibitem{Miller:2007ri}
M.~L.~Miller, K.~Reygers, S.~J.~Sanders and P.~Steinberg,
Ann. Rev. Nucl. Part. Sci. \textbf{57}, 205-243 (2007)
[arXiv:nucl-ex/0701025 [nucl-ex]].

\bibitem{ALICE:2018lyv}
S.~Acharya \textit{et al.} [ALICE],
JHEP \textbf{10}, 174 (2018)
[arXiv:1804.09083 [nucl-ex]].

\bibitem{Guo:2000nz}
X.~f.~Guo and X.~N.~Wang,
Phys. Rev. Lett. \textbf{85}, 3591-3594 (2000)
[arXiv:hep-ph/0005044 [hep-ph]].

\bibitem{Zhang:2003yn}
B.~W.~Zhang and X.~N.~Wang,
Nucl.\ Phys.\ A {\bf 720}, 429 (2003).

\bibitem{Zhang:2003wk}
B.~W.~Zhang, E.~Wang and X.~N.~Wang,
Phys. Rev. Lett. \textbf{93}, 072301 (2004)
[arXiv:nucl-th/0309040 [nucl-th]].

\bibitem{Majumder:2009ge}
A.~Majumder,
Phys. Rev. D \textbf{85}, 014023 (2012)
[arXiv:0912.2987 [nucl-th]].

\bibitem{Wang:1994fx}
X.~N.~Wang, M.~Gyulassy and M.~Plumer,
Phys. Rev. D \textbf{51}, 3436-3446 (1995)
[arXiv:hep-ph/9408344 [hep-ph]].

\bibitem{Zakharov:1996fv}
B.~G.~Zakharov,
JETP Lett. \textbf{63}, 952-957 (1996)
[arXiv:hep-ph/9607440 [hep-ph]].

\bibitem{Chen:2010te}
X.~F.~Chen, C.~Greiner, E.~Wang, X.~N.~Wang and Z.~Xu,
Phys. Rev. C \textbf{81} (2010), 064908
[arXiv:1002.1165 [nucl-th]].

\bibitem{Ma:2018swx}
G.~Y.~Ma, W.~Dai, B.~W.~Zhang and E.~K.~Wang,
Eur. Phys. J. C \textbf{79}, no.6, 518 (2019)
[arXiv:1812.02033 [nucl-th]].

\bibitem{Pang:2016igs}
L.~G.~Pang, H.~Petersen, Q.~Wang and X.~N.~Wang,
Phys. Rev. Lett. \textbf{117} (2016) no.19, 192301
[arXiv:1605.04024 [hep-ph]].

\bibitem{Pang:2018zzo}
L.~G.~Pang, H.~Petersen and X.~N.~Wang,
Phys. Rev. C \textbf{97} (2018) no.6, 064918
[arXiv:1802.04449 [nucl-th]].

\bibitem{Putschke:2019yrg}
J.~H.~Putschke, K.~Kauder, E.~Khalaj, A.~Angerami, S.~A.~Bass, S.~Cao, J.~Coleman, L.~Cunqueiro, T.~Dai and L.~Du, \textit{et al.}
[arXiv:1903.07706 [nucl-th]].

\bibitem{Andersson:1983jt}
B.~Andersson, G.~Gustafson and B.~Soderberg,
Z. Phys. C \textbf{20}, 317 (1983)

\bibitem{Chesler:2007an}
P.~M.~Chesler and L.~G.~Yaffe,
Phys. Rev. Lett. \textbf{99}, 152001 (2007)
[arXiv:0706.0368 [hep-th]].

\bibitem{Casalderrey-Solana:2016jvj}
J.~Casalderrey-Solana, D.~Gulhan, G.~Milhano, D.~Pablos and K.~Rajagopal,
JHEP \textbf{03}, 135 (2017)
[arXiv:1609.05842 [hep-ph]].

\bibitem{Cooper:1974mv}
F.~Cooper and G.~Frye,
Phys. Rev. D \textbf{10}, 186 (1974)

\bibitem{Wang:2023udp}
S.~Wang, Y.~Li, S.~Shen, B.~W.~Zhang and E.~Wang,
[arXiv:2308.14538 [hep-ph]].

\bibitem{ATLAS:2022agz}
G.~Aad \textit{et al.} [ATLAS],
Eur. Phys. J. C \textbf{83} (2023) no.5, 438
[arXiv:2204.13530 [nucl-ex]].

\bibitem{Wang:2021jgm}
S.~Wang, J.~W.~Kang, W.~Dai, B.~W.~Zhang and E.~Wang,
Eur. Phys. J. A \textbf{58} (2022) no.7, 135;
[arXiv:2107.12000 [nucl-th]].

\bibitem{CMS:2019jis}
A.~M.~Sirunyan \textit{et al.} [CMS],
Phys. Rev. Lett. \textbf{125} (2020) no.10, 102001
[arXiv:1911.01461 [hep-ex]].

\bibitem{Vertesi:2024tdv}
R.~Vertesi [ATLAS, CMS, and ALICE],
[arXiv:2405.16955 [nucl-ex]].

\bibitem{Larkoski:2014wba}
A.~J.~Larkoski, S.~Marzani, G.~Soyez and J.~Thaler,
JHEP \textbf{05} (2014), 146
[arXiv:1402.2657 [hep-ph]].

\bibitem{Renk:2012ve}
T.~Renk,
Phys. Rev. C \textbf{88} (2013) no.5, 054902
[arXiv:1212.0646 [hep-ph]].

\bibitem{Brewer:2021hmh}
J.~Brewer, Q.~Brodsky and K.~Rajagopal,
JHEP \textbf{02} (2022), 175
[arXiv:2110.13159 [hep-ph]].

\bibitem{Wang:2024plm}
S.~Wang, Y.~Li, J.~W.~Kang and B.~W.~Zhang,
[arXiv:2408.10924 [hep-ph]].

\bibitem{Kang:2023ycg}
J.~W.~Kang, S.~Wang, L.~Wang and B.~W.~Zhang,
[arXiv:2312.15518 [hep-ph]].

\end{thebibliography}
\end{document}